\documentclass[12pt,preprint]{aastex}

\tighten
\newcommand\be{\begin{equation}}
\newcommand\ee{\end{equation}}
\newcommand\jcd{Christensen-Dalsgaard}

\usepackage{epsfig}
\begin{document}
\shortauthors{Basu and Antia}
\shorttitle{Changes in Solar Dynamics from 1995 to 2002}

\title{Changes in Solar Dynamics from 1995 to 2002}
\author{Sarbani Basu}
\affil{Astronomy Department, Yale University, P. O. Box 208101,
New Haven CT 06520-8101, U.S.A.}
\email{basu@astro.yale.edu}
\and
\author{H. M. Antia}
\affil{Tata Institute of Fundamental Research,
Homi Bhabha Road, Mumbai 400005, India}
\email{antia@tifr.res.in}

\begin{abstract}
Data obtained by the GONG and MDI instruments over the
last seven years are used to study how solar dynamics ---
both rotation and other large scale flows --- have
changed with time.
In addition to the well
known phenomenon of bands of faster and slower rotation
moving towards the equator and pole, we find that the
zonal flow pattern rises upwards with
time. Like the zonal flows, the meridional flows  also
show distinct solar activity related changes.
In particular, the anti-symmetric
component of the meridional flow shows a decrease in speed with activity.
We do not see any significant temporal variations in the dynamics of the
tachocline region where the solar dynamo is believed to be operating.

\end{abstract}


\section{INTRODUCTION}
\label{sec:intro}

Helioseismology has given us the means to study the structure
and dynamics of the solar interior.
The current solar cycle is the first in which we 
can probe the changes taking place within the Sun as the
level of activity changes.
Solar oscillation frequencies are known to change
with time. Elsworth et al.~(1990) studied low-degree
modes ($\ell=0,1,2,3$) of the Sun as obtained by the Birmingham
Solar Oscillations Network (BiSON). 
The Mark-I instrument at the Observatorio del Teide
has yielded low-degree frequencies  for a period of fifteen years between 1984
and 1999. These data too show that the frequencies track solar
activity (Jim\'enez-Reyes et al.~2001).
The first results of solar-cycle related changes in intermediate
degree modes
were reported by Libbrecht \& Woodard (1990).
They found that measurements of solar oscillation frequencies
in 1986 and 1988 showed systematic differences of the
order of 1 part in $10^4$.
With the commissioning of the Global Oscillation Network Group (GONG)
project in 1995 and the Michelson Doppler Imager (MDI) instrument on board the
Solar and Heliospheric Observatory (SOHO) in 1996, we now are able to study the
time variation of solar oscillation frequencies in detail and examine
possible changes taking place in the solar interior.

Changes in solar oscillation frequencies do not indicate any significant changes
in the structure of the Sun as the cycle progresses
(Vorontsov 2001, 2002; Basu \& Antia 2000a, 2001; Basu 2002)
except in the very outer layers. 
We do not yet have reliable frequencies of high degree global
modes that can resolve
changes in the outer layers, however, local-area analyses
around active regions show evidence of structural variations
in the outer layers (Rajaguru, Basu \& Antia~2001;
Bogart, Basu \& Antia~2002). Assuming that the averaged effect of activity on
global mode frequencies is
similar to the effect that local active regions have on modes,
we may expect that if high-degree global modes become available, we shall be
able to resolve the solar-cycle related structural changes in the Sun.
Recently, the mean frequencies of high degree modes have been
reported by
Rhodes et al.~(1998) and
Korzennik, Rabello-Soares \& Schou~(2002), though it is not clear yet if these
will succeed in resolving the near surface layer to identify the
location of solar cycle variation in the solar structure.
Unlike solar structure, solar dynamics  shows significant
changes in the interior. These changes can be important in studying  and modeling
the solar dynamo, which is widely believed to drive the solar cycle.
In this work we shall concentrate on studying changes
in the solar dynamics ---
the rotation rate, as well as other large scale circulations
in the Sun.

Early work has already shown us that the
observed surface differential
rotation of the Sun persists through the convection
zone (CZ) (Christensen-Dalsgaard \& Schou 1988;
Libbrecht 1989; Brown et al.~1989; etc.).
The rotation rate is nearly constant along different
latitudes in most of the CZ. The radiative interior of the
Sun rotates 
almost like a rigid body, with a rotation rate
intermediate between that of the solar equator and pole at the surface
(Thompson et al.~1996; Kosovichev et al.~1997; Schou et al.~1998,
and references therein).
The transition occurs over a fairly thin layer, which
is referred to as the ``tachocline'' (Spiegel \& Zahn 1992).
It is generally believed that the solar dynamo operates in the
tachocline region.
Large-scale circulations within the solar interior are crucial
components of the global dynamo that is generally believed to be
responsible for the Sun's 22-year
magnetic activity cycle. The preferred mechanism for amplifying
the Sun's magnetic field is the generation of a toroidal 
field by shearing a preexisting poloidal field by differential rotation
in the tachocline region (the so called $\Omega$ effect). 
In turn,  it is believed that the 
shear and the overall differential rotation are established by a
combination of Reynolds stresses from the rotationally influenced
turbulent convection and by the associated meridional circulations
(Miesch 2000; Brun \& Toomre 2001).
The poloidal field may be regenerated
from the toroidal field by turbulence (the so called $\alpha$ effect). 
Some new studies of kinetic dynamos suggest that the solar dynamo
mechanism involves not only the above two processes but also an
important third process, the advective transport of the magnetic flux by
meridional circulation as in a conveyor belt 
(see Dikpati \& Gilman 2001; Nandy \& Choudhuri 2002 and references therein).
Dikpati \& Charbonneau (1999) have established a
scaling law between the dynamo cycle period and the meridional flow
speed.  Thus the study of rotation and circulation, and changes therein
are important in our understanding of the causes
of solar variability.

Inversions for the rotation rate show
temporal variations, with bands of faster and slower rotating regions (``zonal
flows'')
moving towards the equator with time (Schou 1999;
Howe et al.~2000a; Antia \&
Basu 2000). These are similar to torsional oscillations observed at the
solar surface (Howard \& LaBonte 1980; LaBonte \& Howard 1982;
Ulrich et al.~1988; Snodgrass 1992).
Torsional oscillations are believed to arise from nonlinear
interactions between magnetic fields and differential rotation.
As such, they should provide a constraint on theories of the solar dynamo.
Covas et al.~(2000) considered an axisymmetric mean field dynamo model
to study temporal variations of the rotation rate and of magnetic fields in
the solar interior. 
They find
temporal variations in the rotation rate that are  similar
to torsional oscillations at low latitudes, and like 
torsional oscillations have bands of faster and
slower rotating regions moving towards the equator with time.
But at high latitudes they find that these bands
migrate pole-wards. As far as surface observations go, some  magnetic features are seen
migrating pole-wards at high latitude (Leroy \& Noens, 1983; Makarov \& Sivaraman 1989).
This high latitude migration is also seen in zonal flows from helioseismic data
(Antia \& Basu 2001; Howe et al.~2001a).
The zonal flows are not just a surface phenomenon, but penetrate at  least
to a radius of $0.9R_\odot$, i.e., to a depth of $0.1R_\odot$ 
(Antia \& Basu~2000; Howe et al.~2000a; Toomre et al.~2000).
But from seismic data it is difficult to obtain the exact depth to which
this flow pattern penetrates, as the errors in inversion increase with
depth and these errors may wipe out the pattern, particularly at high latitudes.
Models of mean field dynamo (Covas et al.~2000, 2001) suggest that these flows
penetrate to the base of the convection zone.
Recent helioseismic studies also show that these flows may
penetrate through most of the convection zone (Vorontsov et al.~2002).
Whether or not there are changes in the dynamics of the tachocline is a 
matter of ongoing controversy.
Howe et al.~(2000b) have reported oscillations with a period of
1.3 years in the solar equatorial region at $r=0.72R_\odot$, but
other studies (Antia \& Basu 2000; Corbard et al.~2001;
Basu \& Antia 2001) did not find any significant changes in the rotation rate
in the tachocline.
The meridional flows also change with time. A preliminary work
by Basu \& Antia (2000b) suggests that there is a correlation between 
solar activity and meridional flow velocities, while Haber et al.~(2001, 2002)
find short time-scale variations in addition to the long-term changes.

In this paper we report the results of a comprehensive study into changes
of solar dynamics. We study the changes in the solar rotation and large
scale circulations, paying particular attention to changes
with solar activity. While it has been usual to look at temporal changes at
different latitudes at the surface or immediately below it, we also study how
the flows change with radial distance at a given latitude to see
if the pattern changes along radius as well.

The rest of this paper is organized as follows: we describe the data and the
analysis technique in \S~\ref{sec:data}; our results are described in
\S~\ref{sec:results}, and we discuss our conclusions in \S~\ref{sec:disc}.

\section{DATA SETS AND TECHNIQUE}
\label{sec:data}

We use data obtained by the GONG and MDI projects for our investigations.
Global modes, which are used to study the
rotation rate, including properties of the tachocline
and the zonal flows, are rather insensitive
to solar meridional flows. To first order in perturbative treatment
the rotational splittings of these modes are not affected by meridional
flows (Woodard 2000).
To study these flows we need to
use local helioseismic methods. In this paper
we use a ring diagram analysis (Hill 1988; Patr\'on et al.~1997) of MDI
data
to study the temporal variation of the meridional flows. 

\subsection{Global modes: data and techniques}
\label{subsec:global}

The solar oscillation frequencies and the splitting coefficients
from the GONG project (Hill et al.~1996) were obtained from 108 day time series. The
MDI data were obtained from 72-day time series (Schou 1999). Some details
of how the GONG and MDI projects determine the frequencies can be found
in Schou et al.~(2002). 
To study the rotation rate, we use data
sets that consist of the mean frequency and the splitting coefficients.
The solar oscillation modes are identified by radial order $n$,
spherical harmonic degree $\ell$ and azimuthal order $m$. For a
non-rotating spherically symmetric star the oscillation frequencies
would be independent of $m$, while rotation and departures from
spherical symmetry lift this degeneracy. The frequencies can
be expressed as
\be
\nu_{n\ell m} = \nu_{n\ell } + \sum_{j=1}^{j_{\rm max}}
a_j(n,\ell ){\cal P}_j^{(\ell )}(m)\;,
\label{eacof}
\ee
where $\nu_{n\ell}$ is the mean frequency of the $(n,\ell)$ multiplet,
$a_j(n,\ell)$ are the splitting coefficients and ${\cal P}_j^{(\ell)}(m)$
are orthogonal polynomials in $m$ (Ritzwoller \& Lavely 1991;
Schou, \jcd\ \& Thompson~1994).
The odd-order splitting coefficients are determined
by the rotation rate in the solar interior and hence these can be used
to determine the rotation rate as a function of latitude and radius.
The splitting coefficients are sensitive only to the north-south symmetric
component of the rotation rate and hence that is the only component that can
be determined with these data.

We use 62 data sets from the GONG
each covering a period of 108 days, starting on
1995, May 7 and ending on 2001, July 21
with a spacing of 36 days between consecutive data sets.
Thus each set, except for the first two and the last two,
overlaps with two others on either side. The data
are from  GONG months 1 to 63, where each GONG `month' represents
an observing period of 36 days.
The MDI data sets (Schou 1999) consist of 28 non-overlapping
sets each obtained from observations taken over a   period of
72 days. The first set begins on 1996 May 1 and the last
set ends on  2002 March 30.

We use a two dimensional Regularized Least
Squares (2D RLS) inversion technique to infer the rotation rate in
the solar interior from each of the available data sets.
The details of the inversion technique have been  described by
Antia, Basu \& Chitre~(1998).
In order to study the temporal variation of the rotation rate we
calculate the rotation rate in the solar interior for each data set
and then take the temporal mean over all data sets at each latitude
and depth. 
This mean can be subtracted from the
rotation rate at any given
epoch to get the time varying component of the rotation rate. Thus we
can write
\begin{equation}
\delta\Omega(r,\theta,t)=\Omega(r,\theta,t)-
\langle \Omega(r,\theta,t)\rangle,
\label{eq:zonal}
\end{equation}
where $\Omega(r,\theta,t)$ is the rotation rate as a function of
radial distance $r$, latitude $\theta$ and time $t$.
Here the angular
brackets denote average over the time duration for which data are available.
This time varying component $\delta\Omega$ is generally called the zonal flow.
We will refer to this time varying component variously  as the residual
in rotation rate or the zonal flow.
The definition of zonal flows is ambiguous: some authors (e.g., Howard \&
LaBonte 1980)
have defined the zonal flows by subtracting the temporal average of
smooth component of rotation rate. The smooth component is generally
defined using three terms (constant, $\sin^2\theta$ and $\sin^4\theta$)
to define the latitudinal variation of rotation rate.
There is some difference in the resulting pattern in the two cases,
as described by Antia \& Basu
(2000). In this work we adopt the definition given by Eq.~\ref{eq:zonal}
where the temporal mean of the full rotation rate is subtracted to obtain
the residuals.

In principle, we can investigate the changes in the tachocline
by subtracting the time-average of the rotation rate at the tachocline
from the rotation rate at each epoch. 
However, the lack of
adequate spatial resolution in the region of the tachocline combined
with the steep radial variation of the rotation rate in that
region  means that inversion results may not be reliable in the
tachocline region.
Thus to determine the properties of the tachocline we use the three techniques
described by Antia et al.~(1998), which are:
(1) a calibration method
in which the  properties at each latitude are determined by direct
comparison with models; (2)
a one dimensional (henceforth, 1d)  method in which  the
parameters defining the tachocline at each latitude  are determined by
a nonlinear least squares minimization using the technique of simulated annealing;
and
(3) a two-dimensional (henceforth, 2d) technique, where the entire
latitude dependence of the tachocline
is fitted simultaneously, again using simulated annealing.
In all techniques the tachocline is represented
by a model of the form (cf., Antia et al.~1998),
\begin{equation}
\Omega_{\rm tac}=
{\delta\Omega_t\over {1+\exp[(r_t-r)/w]}},\label{thi}
\end{equation}
where
$\delta\Omega_t$ is the jump in the rotation rate across the tachocline,
$w$ is the half-width of the transition layer, and $r_t$ the
radial distance of the mid-point of the transition region.
The properties we are interested in are
the position and the thickness of the tachocline and
the change in rotation rate  across the tachocline (i.e., $r_t$, $w$ and
$\delta\Omega_t$). We study these
properties  as a function of latitude and time using all the techniques
listed above.

\subsection{Local-helioseismology: data and technique}
\label{subsec:local}

To study the meridional flow we use three dimensional power spectra obtained
from full disc Dopplergrams from MDI (Bogart et al.~1997).
The Dopplergrams were taken at a
cadence of 1 minute. The area being studied was tracked at the
surface rotation rate (in nHz) given by (Snodgrass 1984)
\be
\Omega(\theta)=451.43 - 54.76\sin^2\theta - 80.17\sin^4\theta.
\label{eq:surf}
\ee
To minimize the effect of foreshortening
we have only used data when the region was passing through the central
meridian.  Each power spectrum was obtained from
a time series of 1664 images covering $15^\circ$ in longitude and
latitude. Successive spectra are
separated by $15^\circ$ in heliographic longitude of the central
meridian. For each longitude, we have used 15 spectra centered at
latitudes ranging from $52.5^\circ$S to $52.5^\circ$N with a spacing
of $7.5^\circ$ in latitude.

We use eight sets of data. In every set we take
average of all available power spectra for each of the 15 latitudes.
The characteristics of the
different sets are listed in Table~\ref{tab:tab1}. This table also lists the
mean 10.7 cm radio flux during the period covered by the data
as obtained from the National Geophysical Data Center web page
at http://www.ngdc.noaa.gov/stp/stp.html.
This quantity is an index of the level of the solar activity during the period
each data set was recorded. With the data used in this work, we can
only look at changes on a time-scale of a few years.
To extract the flow velocities and other mode parameters from the
three dimensional power spectra we fit a model with asymmetric peak profiles as used
by Basu \& Antia (1999), i.e.,
\begin{eqnarray}
&&\hspace{-40 pt}P(k_x,k_y,\nu)= {e^{B_1}\over k^3} +{e^{B_2}\over k^4}+\nonumber \\
&&\hspace{-30 pt}{\exp(A_0+(k-k_0)A_1+A_2({k_x\over k})^2+
A_3{k_xk_y\over k^2})S_x\over x^2+1}\;,
\end{eqnarray}
where
\begin{eqnarray}
x&=&{\nu-ck^p-U_xk_x-U_yk_y\over w_0+w_1(k-k_0)},\\
S_x&=&S^2+(1+Sx)^2,
\end{eqnarray}
and the 13 parameters $A_0, A_1, A_2, A_3, c, p,
U_x, U_y$, $w_0, w_1, S, B_1$ and $B_2$ are determined by fitting the spectra
using a maximum likelihood approach (Anderson, Duvall \& Jefferies~1990).
We fit ridges for each $n$ separately as described by Basu, Antia \& Tripathy~(1999).
The parameter $S$ measures the asymmetry in the peak profile.
The form of asymmetry is the same as that used by Nigam \&
Kosovichev (1998). While all of these parameters may have a 
time variation, in this work we have only considered the variation in flow
velocities determined by $U_x$ and $U_y$.
The fitted $U_x$ and $U_y$ for each mode represents an average of the
velocities in the $x$ and $y$ directions
over the entire region in the horizontal extent and over the vertical
region where the mode is trapped.
We can invert the fitted $U_x$ (or $U_y$) for a set of modes
to infer the variation in
the horizontal flow velocity $u_x$ (or $u_y$) with depth. We have used two different
techniques to invert the velocities, (a) the regularized least squares (RLS)
method, and (b) the method of optimally localized averages (OLA).
The  properties of the inversion results have been describe by
Basu, Antia \& Tripathy (1999).
The velocity component $u_x$ contains information about the
solar rotation rate, modulo the rate at which the region was tracked. The
component $u_y$ is the meridional flow velocity. These results can also
be used to look at the differences between the rotation rate in
the northern and southern hemispheres.

\section{RESULTS}
\label{sec:results}

\subsection{The zonal flows}
\label{subsec:zonal}

\begin{figure} 
\plotone{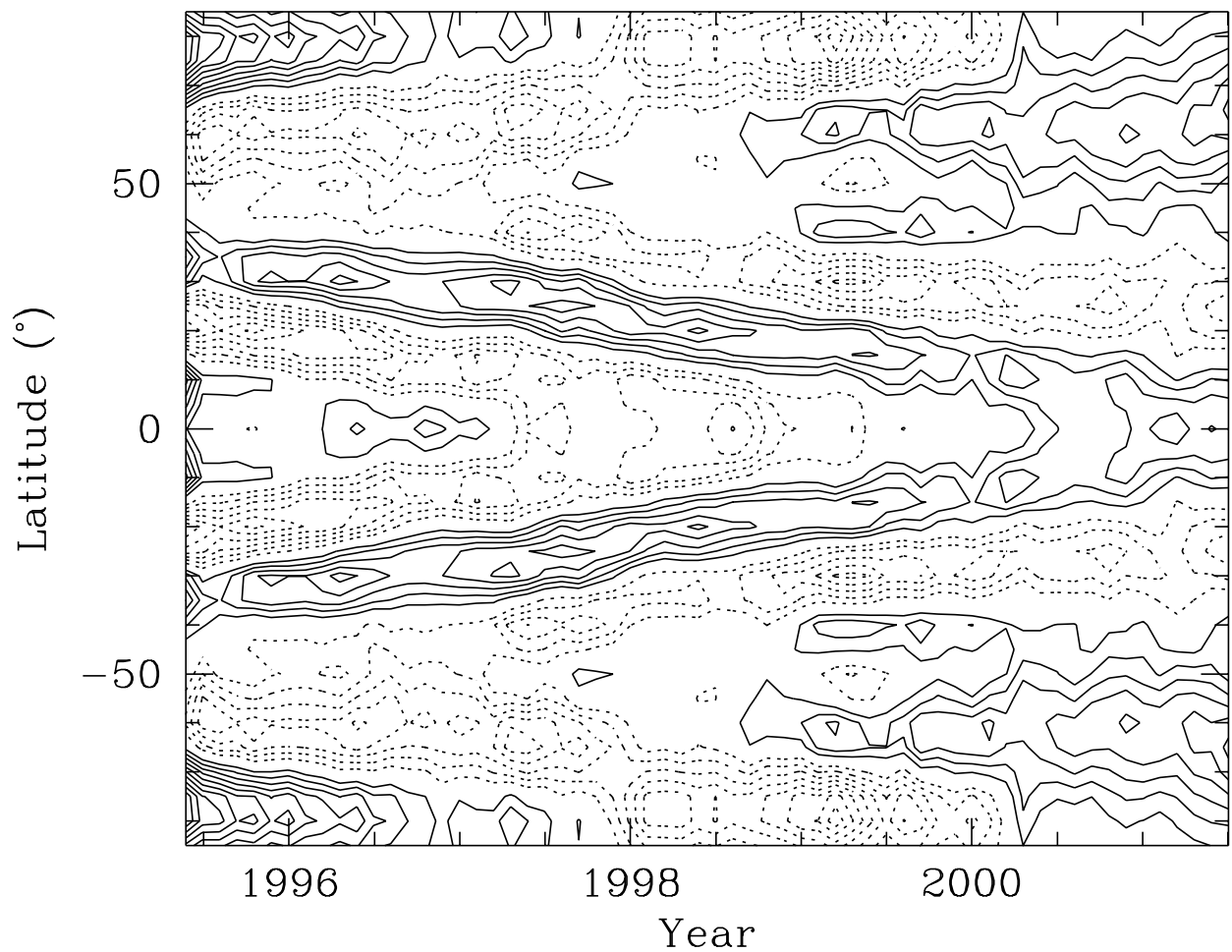}
\caption{Contour diagrams of constant rotation velocity residuals
at $r=0.98R_\odot$ obtained using 2D RLS inversion of
the GONG data. 
The continuous contours are for positive 
$\delta v_\phi$, while dotted contours denote negative values.
The contours are drawn at interval of 1 m/s and the zero contour is
not shown.}
\label{fig:cont}
\end{figure}

The rotation rate in the outer layers of the Sun shows
temporal variations with
a  pattern similar to the well known torsional oscillations at the
surface. To study the temporal variations in the rotation rate we look
at the residuals  obtained
by subtracting the temporal mean of the rotation rate from the rotation rate
at any given  time as explained in Section~\ref{subsec:global} (see Eq.~\ref{eq:zonal}).
The residuals  at $r=0.98R_\odot$
are shown in Fig.~\ref{fig:cont} for the GONG results.
This figure actually shows the linear velocity corresponding to the
residual in rotation rate, i.e.,
$\delta v_\phi=\delta\Omega r\cos\theta$,
where $\theta$ is the latitude.
From this figure it can
be seen that there are distinct bands of faster and slower than
average rotation rate, and that these move towards the equator with time
at low latitudes. This equator-ward movement is well known 
(Schou 1999; Howe et al.~2000a; Antia \& Basu 2000; Vorontsov et al.~2002).
At high latitudes, the bands seem to move towards the poles
as noted by Antia \& Basu (2001) and Ulrich (2001).
The transition between equator-ward and pole-ward movement takes
place around a latitude of $50^\circ$.

\begin{figure}[th]
\plotone{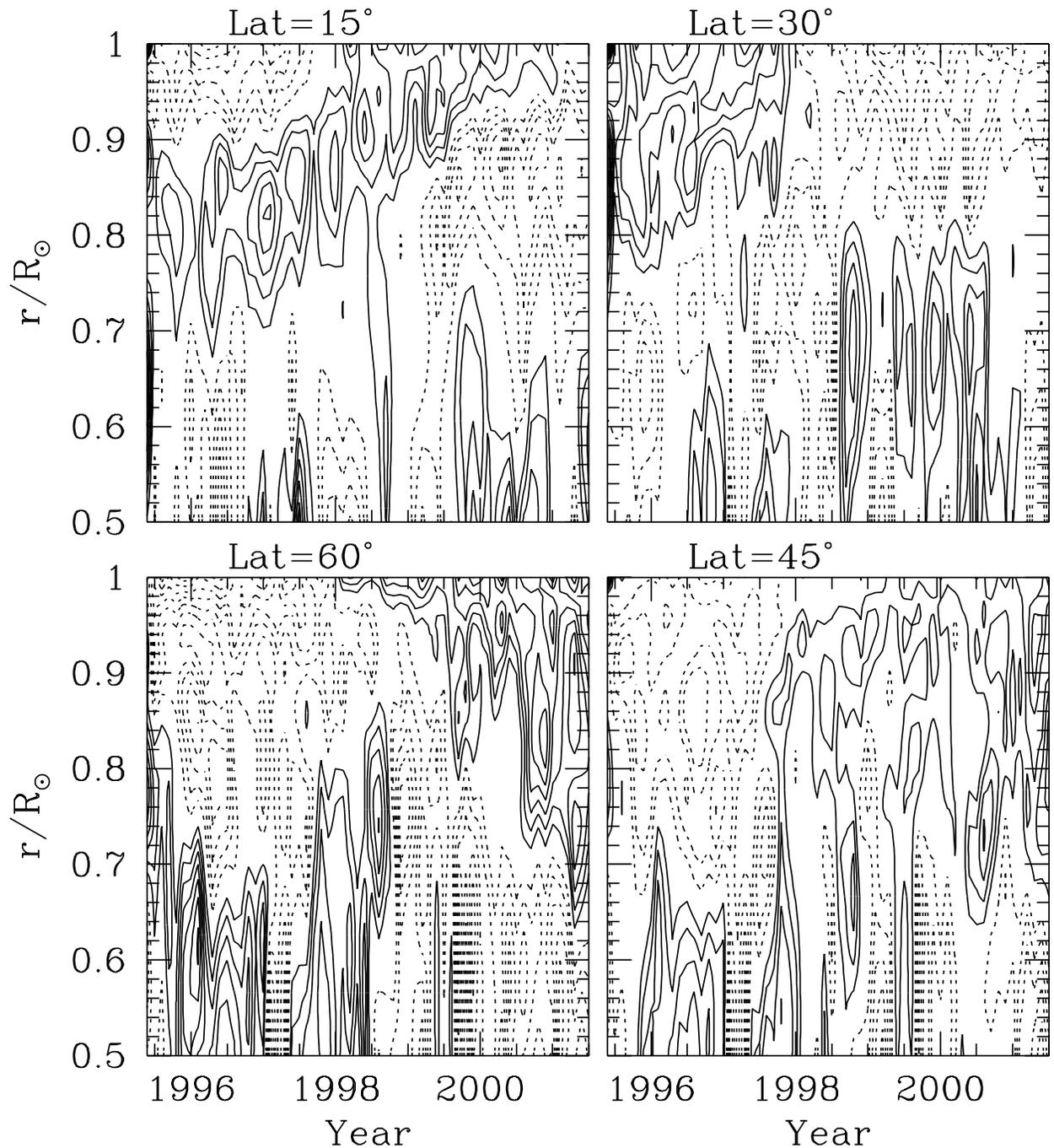}
\caption{The rotation-velocity residuals from the GONG data as a
function of time
and radial distance at selected latitudes. The contours of constant
residual velocity are shown at intervals of 1 m/s, with continuous contours
showing positive values and dotted contours showing negative values.
The zero contour is not shown.}
\label{fig:rotr}
\end{figure}

To study the variation of rotation rate with radial distance
we show  $\delta v_\phi$ obtained from the GONG data plotted as
a function of time and radial distance in
Fig.~\ref{fig:rotr}.
Results for latitudes
of $15^\circ$, $30^\circ$, $45^\circ$ and $60^\circ$ are shown.
At low latitudes there is a clear trend
of contours moving upwards with time. From this
we can deduce that the  pattern
rises upwards with time at a rate of about $0.05R_\odot$ per
year or about 1 m/s, though the upward velocity is probably not
constant.  It should be noted that we
have data covering only a little over half the solar cycle.
The behavior of the zonal flow pattern over the remaining part of
the solar cycle still remains to be seen.
At $15^\circ$ latitude, the pattern clearly
penetrates to radii smaller than $0.9R_\odot$, i.e., to 
depths greater than $0.1R_\odot$ inferred in earlier
works and probably reaches close to the base of the convection zone.
At higher latitudes the depth of penetration cannot be discerned from
these figures. At high latitudes the time evolution of the pattern with
depth is not clear either, if anything it appears that the pattern
goes in the opposite direction, i.e., toward larger depths with
time. At latitude of $60^\circ$ again the band of faster rotation
appears to penetrate nearly to the base of the convection zone near
the maximum of solar activity. This is probably similar to the pattern
seen by Vorontsov et al.~(2002).
This change in behavior (sinking rather than a rising pattern)
seems to occur at the latitude of $50^\circ$ where the zonal
flow pattern changes
from one moving towards the equator to one moving towards the
pole. It is possible that at all latitudes during part of the cycle
the pattern sinks downwards. Only data over complete cycle can
show if this is indeed the case.

\begin{figure}
\epsscale{.60}
\plotone{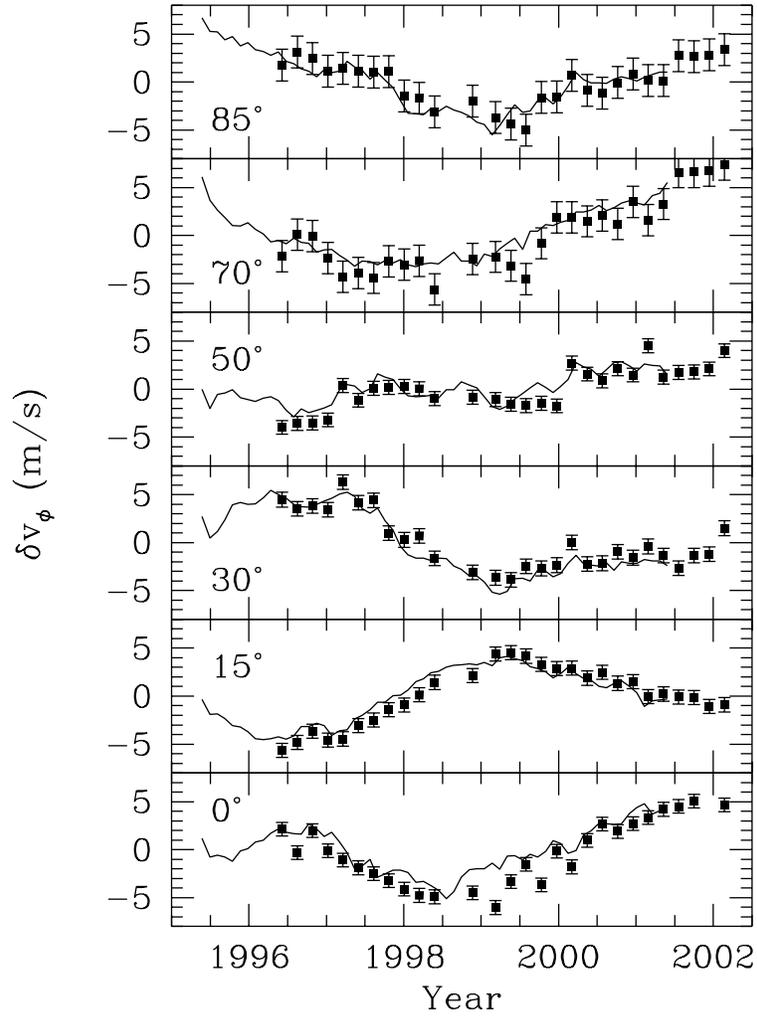}
\caption{The zonal flow velocity as a function of time at different 
latitudes at $r=0.98R_\odot$. The latitudes are marked in each panel.
The solid line shows the GONG results, while points with error bars
show the MDI results.
}
\label{fig:lat}
\end{figure}

The changing pattern of the zonal flows implies that the
maximum and minimum velocities for each latitude occur at 
different times (Schou 1999, Howe et al.~2000a). This is seen in Fig.~\ref{fig:lat} which
shows the zonal flow velocity at different latitudes
as a function of time at $r=0.98R_\odot$ from both the GONG and MDI data.
It may be noted that the temporal average is taken over different
time intervals for the GONG and MDI data, which can introduce a systematic
shift in some cases when residuals are calculated. This appears to be
the case at the equator.
We see that at any given time, different latitudes are at different phases
of the pattern.
The location of minima and maxima have strong
latitudinal variations. At latitudes of around $50^\circ$
the amplitude
of variation is distinctly smaller as compared to other latitudes.
This is close to the latitude where there is a switch from equator-ward
movement to pole-ward movement.
While the periodicity of this pattern can only be confirmed with more
data covering a few solar cycles, from the results over limited
duration it is clear that 
the pattern is not sinusoidal. This might
explain the higher harmonics of the solar cycle that 
Vorontsov et al.~(2002) find.

\begin{figure} 
\plottwo{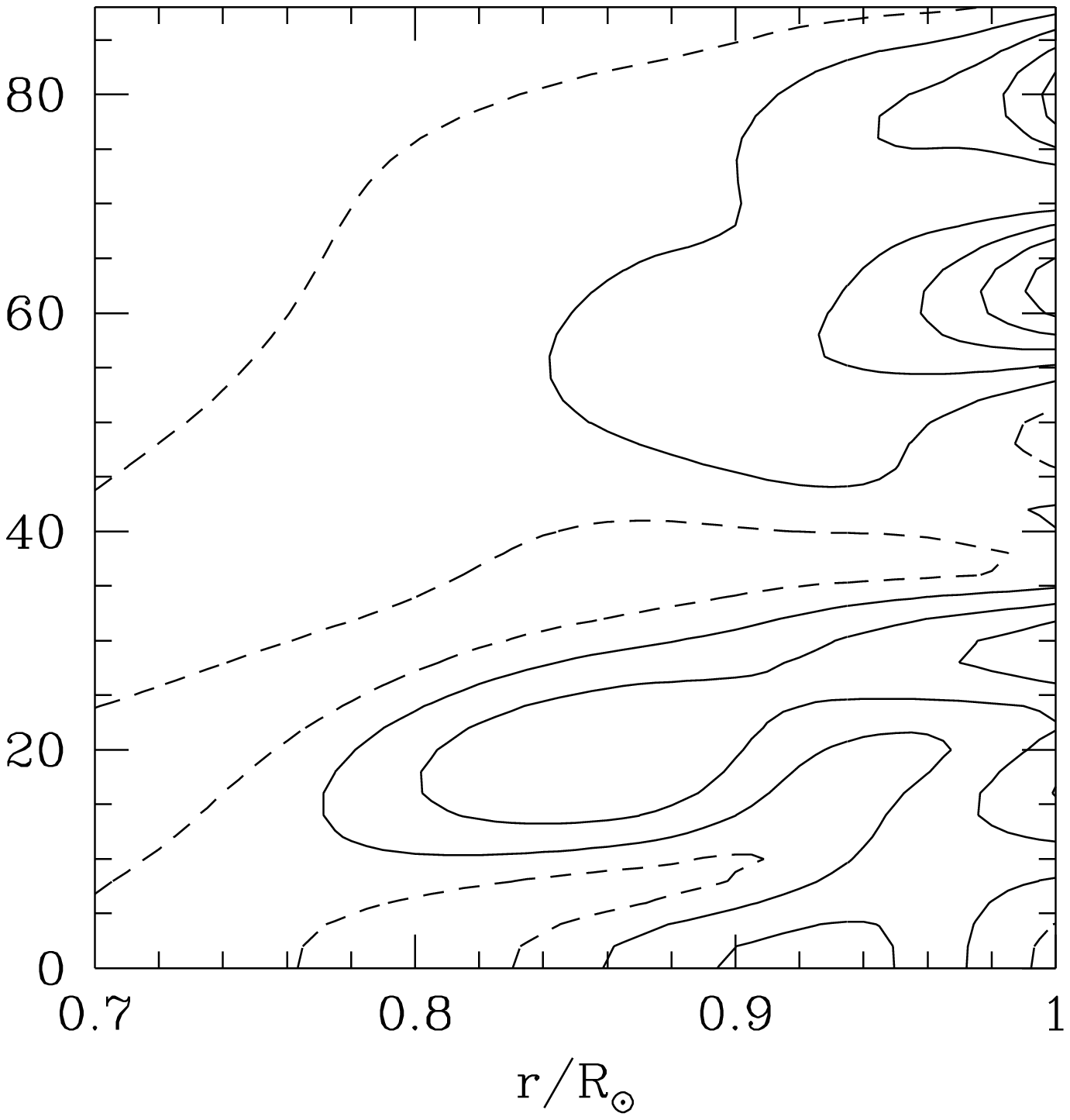}{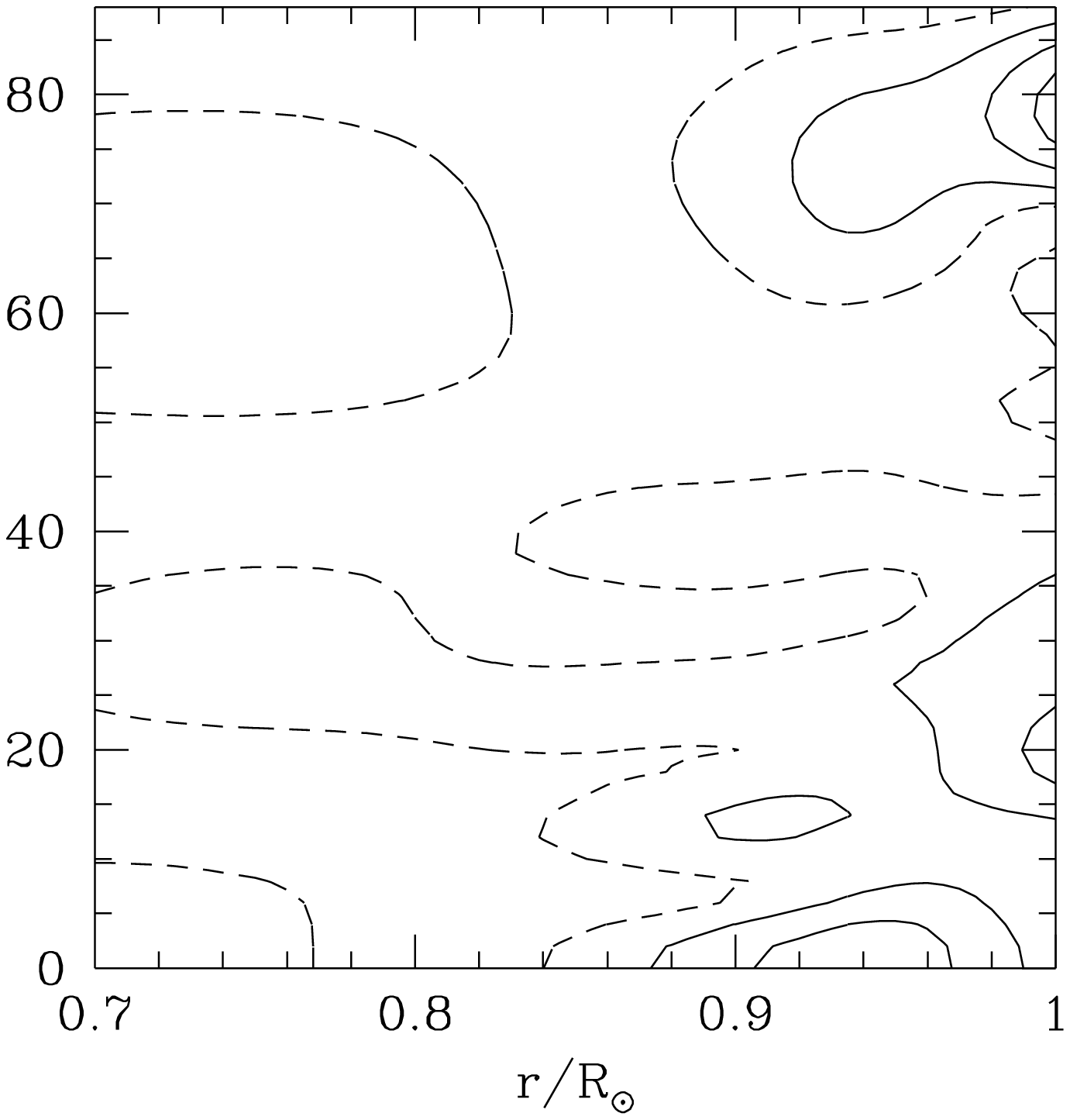}
\caption{Contour diagram showing amplitudes of the $k=1$ and $k=3$
(Eq.~\ref{eq:harm})
components of zonal flow expansion as obtained using
GONG data,  assuming a period of 11 years. 
The contours are at intervals of 1 m/s with dashed contour representing
the value of 1 m/s. The left panel shows $a_1$
while the right panel shows $a_3$.}
\label{fig:harm}
\end{figure}

It is not clear if  it is
possible to discern the periodicity in rotation rate at various
latitudes and depths from data for a limited time interval.
But if we assume that the period is 11 years,
we can try to fit an expansion of the form
\be
\delta v_\phi(r,\theta,t)=\sum_{k=1}^N a_k(r,\theta)\sin(k\omega t+\phi_k),
\label{eq:harm}
\ee
where $\omega=2\pi/11$ yr$^{-1}$ is the basic frequency of the solar cycle.
Here $a_k$ is the amplitude of $k$th harmonic and $\phi_k$ its phase.
Since we have data over only a part of the solar cycle, different
components in Eq.~\ref{eq:harm} are not orthogonal and their
relative amplitudes depend on which components are included
in the fit. We find that inclusion of $k=2$ component tends to
suppress the amplitude of the fundamental ($k=1$) component,
while inclusion of the $k=3$ component does not affect the amplitude of
$k=1$ component. 
To illustrate the
results Fig.~\ref{fig:harm} shows  contour diagrams for the amplitudes
of the $k=1$ and 3 components, when only components for $k=1,3,5$ are
used in the fit. The first component is qualitatively
similar to that found by Vorontsov et al.~(2002) who also find
a prominent feature
extending through most of the convection zone at high latitudes.
At middle latitudes of around $45^\circ$ the amplitude is small.
At low latitudes again it is clear that the pattern penetrates
to depths close to the base of the convection zone as inferred
above from the study of radial behavior of the zonal flow velocities.
It is clear that the $k=3$ harmonic is significant, though its
magnitude is about half of the first term.
Vorontsov et al.~(2002) also find the third harmonic in their
analysis.
The significant amplitude of higher components may be expected from
the non-sinusoidal form of the temporal variation. It could also
arise if the basic period is different from 11 years. From the limited
data that we have,  we cannot distinguish between these possibilities.
The relative magnitude of various harmonics will depend on the origin
of the zonal flow pattern. 
Below the convection zone the amplitude
of the main component is very small, indicating that solar cycle
variations are not seen in these layers. 
At very deep layers, close to the tachocline, the errors
on the zonal flow velocities are so large, that we 
cannot assign any significance to the fluctuating results
obtained there. 

\subsection{The tachocline}

\begin{figure} 
\plotone{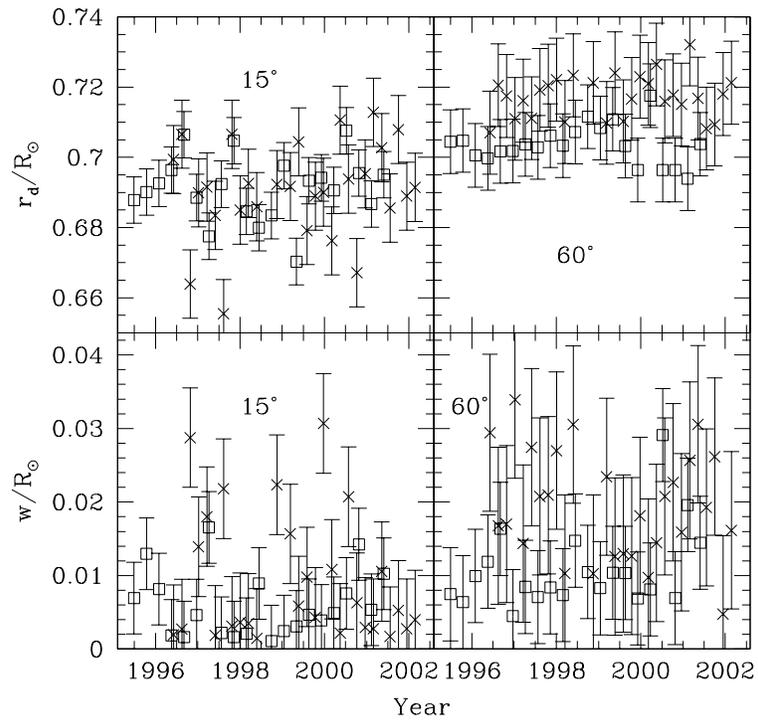}
\caption{The mean radial position and half-width 
of the tachocline at a latitude of $15^\circ$ and $60^\circ$
is shown as a function of time for the GONG (crosses) and MDI (squares)
data. These results are weighted averages of those obtained using
the three techniques mentioned in \S2.1.}
\label{fig:tach2}
\end{figure}

To study possible temporal variations in the properties of the
tachocline, we use each of the non-overlapping data sets from
the GONG and MDI to determine the mean location, width and the jump
in the  rotation rate across the tachocline. To improve the precision
of the results,
we take the weighted mean of the values obtained using the three methods
mentioned in \S \ref{subsec:global}. Each result is weighted inversely
with the square of the estimated error while calculating the mean value.
Fig.~\ref{fig:tach2} shows the
position and the half-width of the tachocline
as a function of time for both the GONG and MDI data.
Note that while taking averages the errors have been
estimated by assuming that there is no correlation
between different estimates. This may, in principle, cause the
errors to be underestimated. However, looking at the scatter between
different points in Fig.~\ref{fig:tach2} it appears that errors
may actually be overestimated, particularly at high latitudes.
This could be due to correlation between different tachocline
parameters in our fits.
It can be seen that none of these properties show any significant
temporal variation. There appears to be some systematic difference
between  GONG and MDI results for depth of the tachocline at high
latitude. This could be due to known differences between the two
data sets (e.g., Howe et al.~2001b; Schou et al.~2002) which mainly arise from differences
in the data reduction techniques.
The variation in
$\delta\Omega_t$ is not shown in the figure, but that also does not
show any clear temporal variations.

Since we do not find any significant temporal variations in the properties of
the tachocline, we can take the temporal average of all sets to
improve the precision of the results,  and the averages at a few 
latitudes are shown in Table~2. These averages are obtained by
combining all non-overlapping sets in both the GONG and MDI data.
From this table it is clear that there is a distinct latitudinal
variation in the position and thickness of the tachocline. These
results are consistent with those of Charbonneau et al.~(1999) and
Basu \& Antia (2001). However, it appears that
the latitudinal variations in the position and width may not be continuous.
The position and width at latitudes of $0^\circ$ and $15^\circ$ are
the same within error bars. Similarly, those at latitudes of
$45^\circ$ and $60^\circ$ are close to each other. In between these pairs of
latitudes, is the transition in the sign of $\delta\Omega_t$, the
jump in rotation rate across the tachocline.
At low latitudes the rotation rate
increases with radial distance, while at high latitudes the
rotation rate  decreases with radial distance in the tachocline.
Thus it is possible that the tachocline consists of two parts one at
low latitudes and the other at high latitudes which are located at
different depths and have different widths, but there may be no variation
in position and width within each of these regions. Fig.~\ref{fig:tach}
shows such a  representation of the tachocline. In this figure we have used
a half-width of $2.5w$ to cover the region where $85\%$ of the
transition in rotation rate is complete. This is equivalent to the
width of the tachocline as defined by Kosovichev (1996) and Charbonneau
et al.~(1999). At the base of the convection zone the latitudinal
resolution of observed frequency splittings is limited and from these
results it is not possible to decide if there is indeed a discontinuity
in the tachocline properties as a function of latitude or whether the
variation is gradual between latitudes of $15^\circ$ and $45^\circ$.
If the variation is confined to a very narrow latitude range than
for all practical purposes it can be considered as a discontinuity.
Clearly, more work is needed to confirm whether the latitudinal variation is indeed
steep around a latitude of $30^\circ$.

\begin{figure} 
\plotone{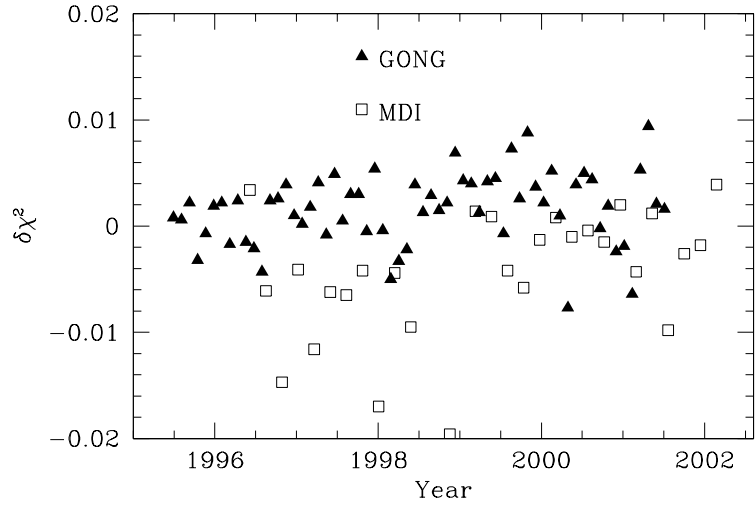}
\caption{The difference in $\chi^2$ per degree
of freedom between 2d fits to the  continuous functions and those
to functions with discontinuity in latitude.}
\label{fig:chi}
\end{figure}

\begin{figure} 
\plotone{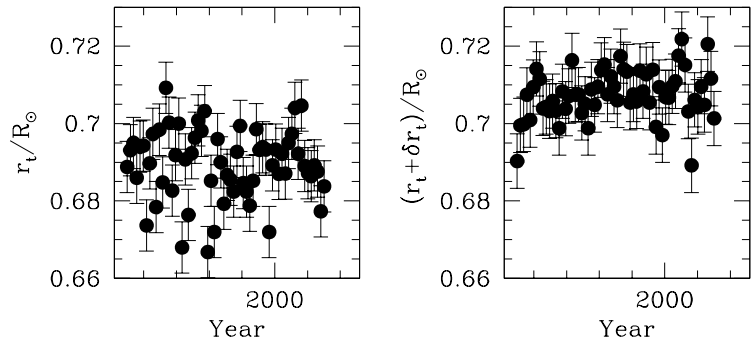}
\caption{The tachocline position
as obtained from 2d fits to a function with discontinuity in $\theta$
to the GONG data.
The left panel shows the results at low latitude while the right panel
shows the results for high latitudes.}
\label{fig:dis}
\end{figure}

In order to test if there is indeed a discontinuity in the position
and width of the tachocline as a function of latitude, we repeat
the 2d simulated annealing fit to the tachocline properties assuming a discontinuity
in $r_d$ and $w$ at $\theta=30^\circ$. The difference in $\chi^2$
per degree of freedom for fits with discontinuity and those with continuous
variation with $\theta$ is shown as a function of time in
Fig.~\ref{fig:chi}. The GONG data appears to favor a discontinuity, though the
difference in $\chi^2$ is rather small.
The MDI data do not show any clear difference.
On average for  MDI results, the $\chi^2$ increases in the presence of 
discontinuity, and thus these results
do not favor discontinuity in latitudinal dependence,
but the scatter in MDI points is much larger than those in GONG
points. Moreover, there appears to be some systematic increase in
$\delta\chi^2$ between the pre- and post-recovery data sets from MDI.
It is not clear if this is due to possible systematic errors due
to variations in MDI characteristics
during recovery (Antia 2002) or a temporal variation or simply
statistical fluctuation.
We do not know why this difference in behavior is seen between the
GONG and MDI results.
The GONG data also show a distinct temporal variation in
the $\chi^2$ difference, thus suggesting that there may be a
temporal variation in the magnitude of the discontinuity at $\theta=30^\circ$.
Fig.~\ref{fig:dis} shows the position of the tachocline in the two regions as a
function of time as obtained using the GONG data.
There is a  very marginal variation in the position
of the tachocline at high latitudes, which is within error bars.
This
figure can be compared with Fig.~\ref{fig:tach2} which shows the same results
using other techniques.
Note that in both Figs.~\ref{fig:tach2} and \ref{fig:dis} the scatter
in position is much less at high latitudes. In Fig.~\ref{fig:dis} this
is clearly due to the correlation between $r_t$ and $\delta r_t$ which
can be verified by looking at the fits to these two parameters.
A similar correlation is present in the fits to the continuous
function too,  resulting in smaller scatter at high latitudes, which leads to
an overestimate of errors.  It is
not clear if this temporal variation is significant. If this variation is real,
it will imply that the high latitude part of the tachocline moves
up or down with the solar activity and this movement may play a crucial
role in the operation of dynamo. The corresponding variation in
the width is not clear at all.

\subsection{North-South asymmetry in the rotation rate}

To study the north-south antisymmetric component of the rotation rate
we use the ring diagram technique.
The inverted values of $u_x$ contain information about the 
rotation rate in the solar interior.
Since the rate at which the regions were tracked is symmetric
about the equator, any difference between the inverted $u_x$ at a given
latitude of the northern hemisphere and that at the same latitude in the
southern hemisphere is caused by the asymmetry of the rotation rate
about the equator.

\begin{figure} 
\plotone{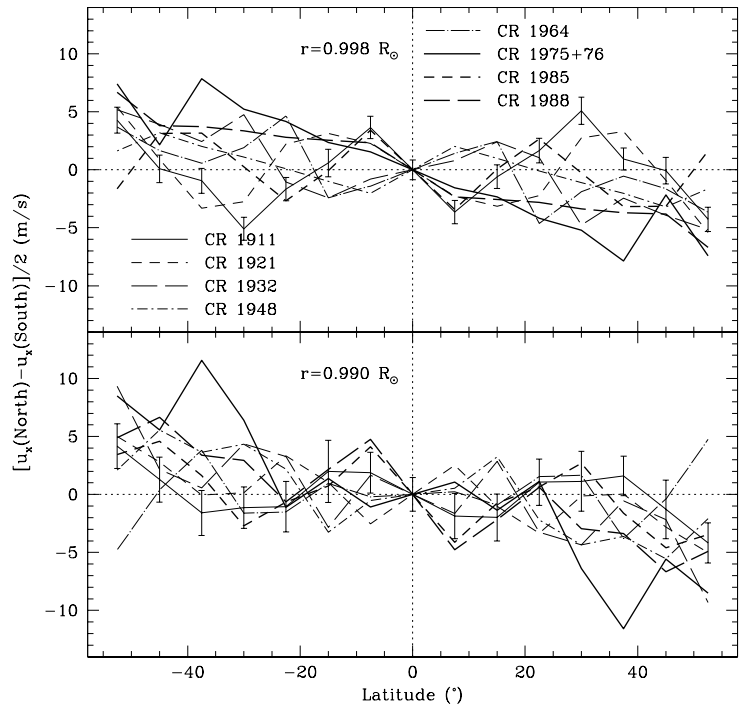}
\caption{The North-South antisymmetric component of the
solar rotation velocity as
a function of latitude at two different radii.
Only OLA results are shown.
RLS results are very similar. These results were
obtained by the ring-diagram analysis of MDI data.
}
\label{fig:asym}
\end{figure}

The north-south antisymmetric component of the rotation velocity at various times are
shown in Fig.~\ref{fig:asym} at two sample radii. The component  appears to change somewhat with time, however,
it is not
clear if there is any systematic temporal variation. It is possible that
we have to look at intermediate time-steps to discern any pattern.
Alternatively, it is possible that the apparent variations are 
caused by errors in pointing or tracking, and/or errors in inversions.
In the outer layers, above a radius of about 
 $0.98R_\odot$, the antisymmetric component appears to be
significant. An average over all data sets  shows a significant
slope in these layers. Below this depth the errors in inferred
velocity are too large to make any definite conclusion.

\subsection{The meridional flow}

\begin{figure} 
\plotone{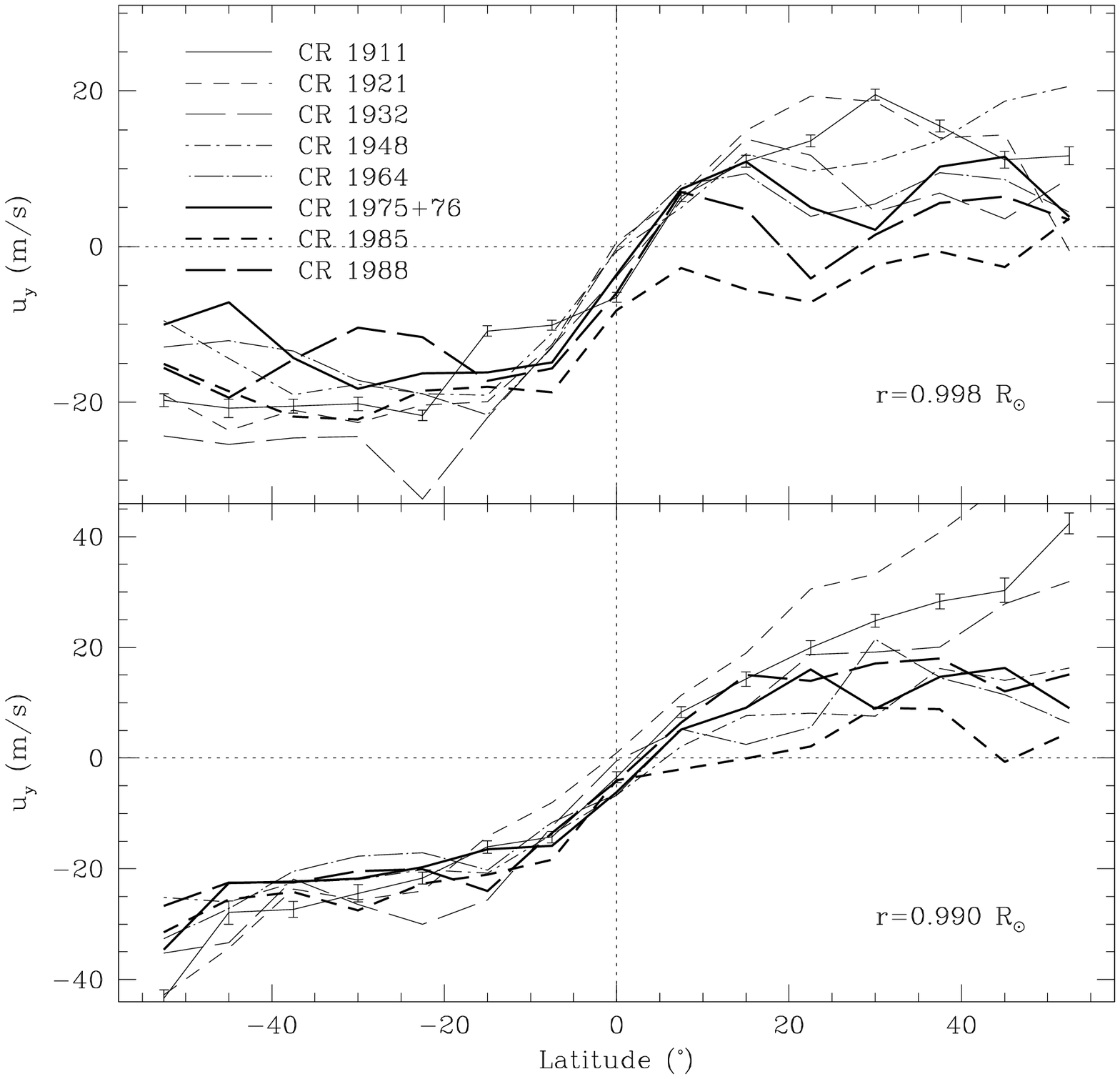}
\caption{The meridional flow velocities
obtained from the different data sets plotted as a function of
latitude for two different radii. Only OLA results are shown.
RLS results are very similar.
}
\label{fig:merid}
\end{figure}

\begin{figure} 
\plotone{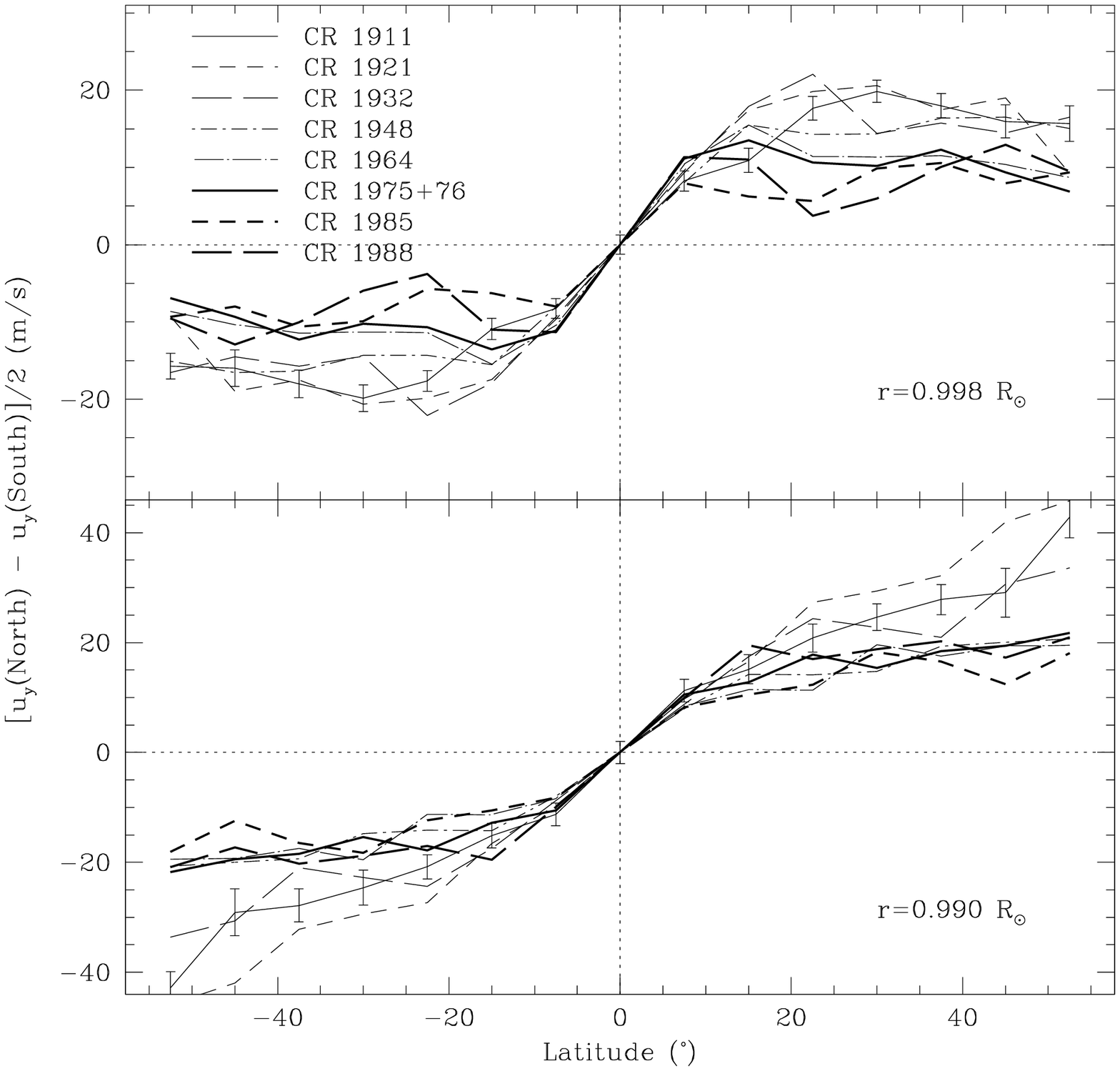}
\caption{The north-south antisymmetric component of meridional flow velocities
obtained from the different data sets plotted as a function of
latitude for two different radii. Only OLA results are shown.
RLS results are very similar.
}
\label{fig:merid_asym}
\end{figure}

The meridional flow velocity, $u_y$, at two representative
depths  is shown as a function of latitude in Fig.~\ref{fig:merid}.
The results of all eight data sets are shown.
The dominant component in meridional velocity is the north-south
anti-symmetric component with a variation  of form $\sin(2\theta)$.
In both hemispheres the meridional flow is from equator towards
the poles.
The meridional flow shows a definite, and systematic
time variation. The maximum velocity of the flow is smaller
when the Sun is more active.
There is a small southward flow at the equator 
at all times,  but that could be an
artifact of pointing errors or errors in the Carrington elements
used to define the rotation axis or other systematic errors in
the ring diagram analysis, and as such its significance 
is difficult to ascertain. This southward
flow was also seen by Giles et al.~(1997) and Basu et al.~(1999).
We note that the expected `S' shape
of the flow is seen only at very shallow depths. At deeper layers,
the data sets from low activity period do not show any tendency
of lowering velocity at high latitudes, while the data sets during
high activity period do.  The time variation is
seen more clearly if we plot the north-south antisymmetric
component of the flow, which is shown in Fig.~\ref{fig:merid_asym}.
In this figure the decrease in velocity at high latitudes with time
can be clearly seen. 
The flow speeds at low latitudes show almost no change with time, the
intermediate and high latitudes do.

\begin{figure} 
\plotone{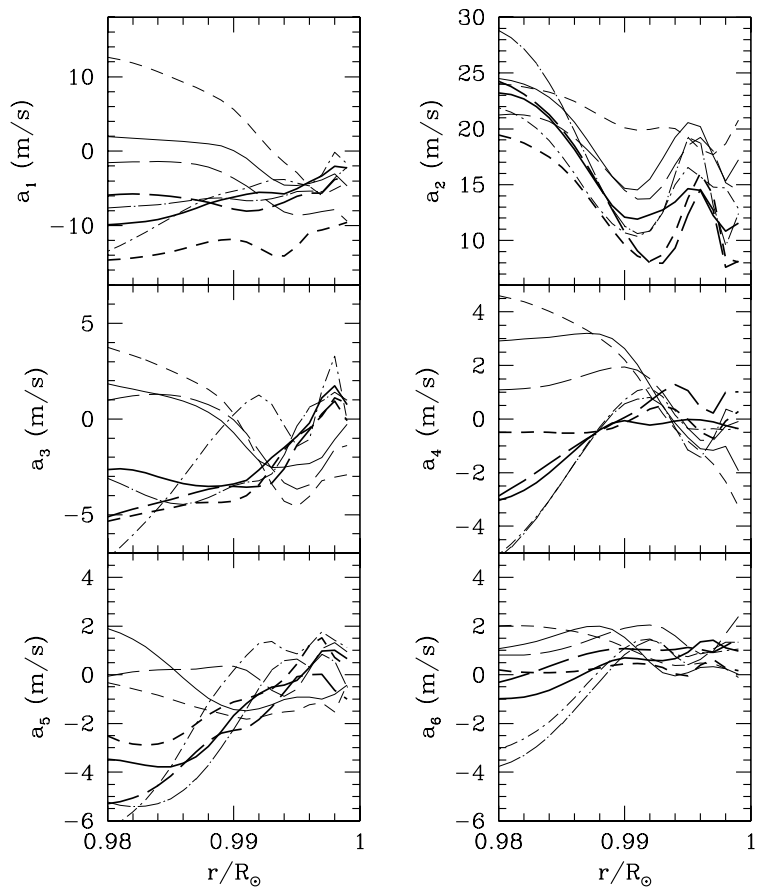}
\caption{The amplitude of the different components
of the meridional flow as defined in Eq.~(\ref{eq:comp}) plotted as a
function of radius.
The different line types represent the
different data sets (see legend of Figs.~\ref{fig:merid} and
\ref{fig:merid_asym}). Only RLS results are shown, OLA
results are very similar. The error bars are not shown for clarity,
but these can be seen in Fig.~\ref{fig:comp_time}.
}
\label{fig:comp}
\end{figure}

\begin{figure}
\plotone{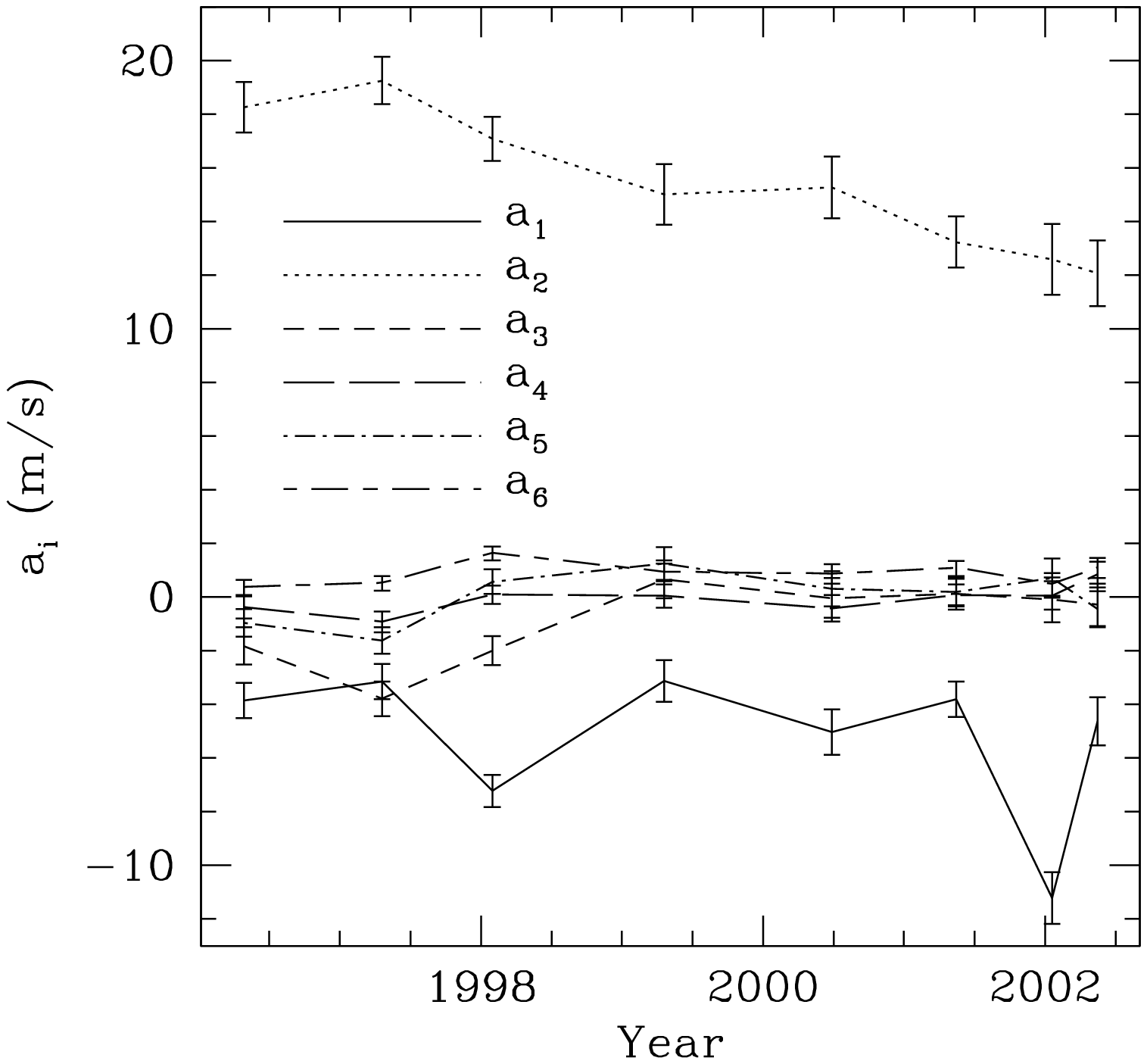}
\caption{The amplitudes of the different components of the
meridional flow plotted
as a function of time at $r=0.995R_\odot$.
}
\label{fig:comp_time}
\end{figure}

To take a more detailed look at the meridional
flow variations we decompose the flow velocity into different components
(see Hathaway et al.~1996):
\begin{equation}
u_y(r,\phi)=-\sum_i a_i(r)P^1_i(\cos(\phi)),
\label{eq:comp}
\end{equation}
where $\phi$ is the colatitude, $P^1_i(x)$ are the associated Legendre
functions. The first six components are
found to be significant and are shown in Fig.~\ref{fig:comp}.
These components also show a variation with time.
For most of these components the temporal variation is not 
systematic, and could be due to errors in estimating their amplitudes.
We plot the amplitudes of the different components as a function
of time at a sample radius of  $r=0.995R_\odot$ in Fig.~\ref{fig:comp_time}.
We can see clearly that only the amplitude of the 
dominant  component, $a_2$, shows a systematic
variation with time, with its magnitude reducing
with time during the time period covered by the MDI data.
The north-south symmetric component $a_1$ is not zero,
and always shows a south-wards flow. It does show a 
change with time, but it is difficult to say whether it
is real, since a changing pointing-error would give
a similar result. The symmetric component
$a_3$ is also non-zero, but again its significance is
hard to judge. The higher order components do not seem to
show significant temporal variation.
Other than $a_2$ and perhaps $a_1$, no component seems to
show  a definite change with time either. Similar results are found
at radii larger than $0.99R_\odot$,
but in deeper layer no systematic
temporal variation is seen in any of these components.

\section{DISCUSSION AND CONCLUSIONS}
\label{sec:disc}

We have done a comprehensive analysis of the available helioseismic
data from the GONG and MDI projects to study changes in the solar
dynamics with changes in the level of solar activity.
The rotation-rate residuals obtained by subtracting the time-averaged
rotation rate from that at each epoch show the well-known 
pattern of zonal flows similar to the torsional oscillations observed
at the surface. This pattern consists of bands of faster and
slower than average rotation rate moving towards the equator with time
at low latitudes. At high latitudes these bands move towards the
poles. The transition between equator-ward and pole-ward movement
occurs around a latitude of $50^\circ$. Observations of magnetic
features at the solar surface also show similar pole-ward movement
at high latitudes (cf., Leroy \& Noens 1983; Makarov \& Sivaraman 1989;
Erofeev \& Erofeeva 2000; Benevolenskaya et al.~2001).
Theoretical models of Covas et al.~(2000, 2001) using a mean-field
dynamo also show this feature.

The zonal flow pattern at low latitudes appears to be rising upwards
with time at a rate of about 1 m/s at low latitudes. A similar conclusion was drawn
by Komm, Howard \& Harvey~(1993) by comparing the zonal flow pattern
obtained using the Doppler
measurements at the surface with those from magnetic features which
are believed to be anchored below the surface.
Furthermore, from the
variation with depth it appears that the zonal flows penetrate
through a good fraction of the convection zone. This appears to
contradict earlier inferences (Howe et al.~2000a; Antia \& Basu 2000)
that these flows penetrate to a radius of $0.9R_\odot$ only.
This difference may be because the errors in inversion increase
with depth thus making it difficult to decide whether the pattern
continues to deeper layers. Particularly, at high latitudes the
errors are large and it may be difficult to see the pattern in
deeper layers. In these earlier works an attempt was made to look
at the pattern at a range of latitudes and that may be the reason
that it was difficult to verify if it penetrates through the convection
zone. If we look only at low latitudes where errors are comparatively
small, it appears that the pattern penetrates deeper.
Recently, Vorontsov et al.~(2002) also found that
a band of faster rotating elements appear to penetrate almost up to the
base of the convection zone at latitude of around $60^\circ$.
We also find similar results at this latitude. The mean-field
dynamo models of Covas et al.~(2000, 2001) also predict that the zonal
flow pattern should persist through the convection zone. Our
results suggest that this is likely to be true, though at high
latitudes the situation is not clear. More data covering the
entire solar cycle may be able to resolve this issue.

\begin{figure} 
\plotone{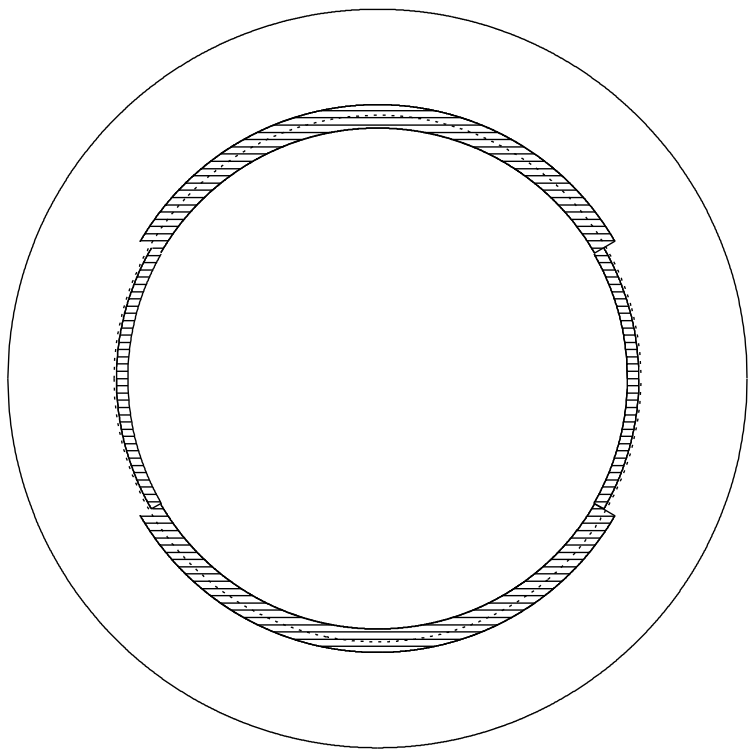}
\caption{The location of the tachocline inside the Sun. The shaded area
is the tachocline with a half-width of $2.5w$. The dashed line shows
the base of the convection zone at $r=0.713R_\odot$
(\jcd, Gough \& Thompson 1991; Basu \& Antia 1997).}
\label{fig:tach}
\end{figure}

It is conventionally assumed that the solar dynamo is located in the
tachocline region below the base of the convection zone, and hence,
one would expect changes there. However,
we do not find any significant temporal variation in the properties
of the tachocline. A temporal mean over all data though shows
a clear latitudinal variation in the depth and thickness of the
tachocline. Similar latitudinal variations have been seen earlier
(Charbonneau et al.~1999; Basu \& Antia 2001).
A closer look at this variation suggests that the tachocline
may consist of two parts, one at low latitudes where the rotation
rate increases with radial distance and other at high latitudes
where the rotation rate decreases with $r$. These parts appear to
have different depth and thickness, but there may not be much
latitudinal variation within each of these parts. On the other
hand, it is known that there is no significant latitudinal variation in the
depth of the convection zone (Basu \& Antia 2001). Similarly,
any possible latitudinal variation in sound speed at the
base of the convection zone is less than $10^{-4}$ (Antia et al.~2001).
The latitudinal variation in the position of the tachocline
is however highly significant, being $(0.019\pm 0.003)R_\odot$.
 It is
generally believed that shear within the tachocline gives rise to
some mixing which accounts for the low lithium abundance on the solar
surface as well as the bump in sound speed difference seen in solar
models
which do not include this mixing (\jcd\ et al.~1996).
Inclusion of mixing in the tachocline removes the bump
(Richard et al.~1996; Brun, Turck-Chi\`eze \& Zahn~1999).
Thus if the position and thickness of the tachocline depend on the
latitude, the extent of mixing should also depend on the latitude
which should give rise to a strong latitudinal variation in the solar structure.
However, we do not see any evidence for significant latitudinal variation in the solar
structure. This would imply that the tachocline cannot be responsible
for mixing, unless the depth and thickness of the tachocline vary
in such a manner that the lower boundary of the mixed region is
essentially independent of the latitude. From Fig.~\ref{fig:tach} it can be
seen that this is indeed the case, with the lower boundary of the
tachocline being essentially independent of  latitude. Of course,
the thickness of $2.5w$ is chosen to achieve this purpose, but it
is a reasonable estimate of the mixing region. Thus if the thickness
of the tachocline varies along with the depth in a suitable manner
it may be possible to ensure that the solar structure is close to
being spherically symmetric. From this point of view also it is
simpler to assume that the tachocline consists of two parts,
rather than assuming that it has a 
continuous latitudinal variation. In the
latter case one will require the thickness to match the latitudinal
variation in depth at every latitude.

It may be argued that we will need some coincidence to match the
variation in depth and latitude even with a two component model for
the tachocline. It may be noted that in the absence of
any mixing, the squared sound speed in a solar model departs by about
0.5\% from the solar values as inferred by inversions
(Christensen-Dalsgaard et al.~1996) while the latitudinal variation
in the sound speed in the tachocline region is less than
a few parts in $10^5$. It is not clear what
causes this agreement between depth and thickness variations.
From Fig.~\ref{fig:tach} it appears that
at low latitudes the upper limit of the tachocline is below the
base of the convection zone and  hence there could be some region
which is not fully mixed. This difference is very small and could be a
result of errors in the deduced values of the
tachocline position or thickness.  A very
thin layer that is not fully mixed may not result in any significant
anomaly in the solar structure.

To test the two component hypothesis for the tachocline, we attempt
a 2d simulated annealing fit with discontinuity in latitude for position and
width. The fits to the GONG data show a marginal reduction in $\chi^2$
per degree of freedom when discontinuity is introduced in latitudinal
variation. It may be noted that the number of parameters to be fitted
is the same in both continuous and discontinuous case. This result
is not confirmed by the MDI data which do not show any systematic differences
in $\chi^2$. The reason for this discrepancy is not understood. The
GONG data also indicate that the position of the tachocline at
high latitude may be moving upward with activity. The temporal
variation is less than $1\sigma$ error bars and its significance
is not clear.

The north-south antisymmetric component of rotation rate cannot
be detected from the splittings of global modes. We use the
ring diagram technique to study this component in the outer convection
zone. We do not find any definite temporal variation in this
component. Averaging over time we find this component to be
significant above $r=0.98R_\odot$, where it has a magnitude of
about 5 m/s at a latitude of $45^\circ$.
It is quite possible that this is an artifact of systematic errors
in ring diagram analysis.

The meridional component of velocity can also be studied using
the ring diagram technique. The meridional flow velocity shows
a clear temporal variation as its magnitude seems to decrease with
increase in the solar activity. The dominant component of the meridional
flow velocity, i.e., $1.5a_2\sin(2\theta)$ shows a systematic
decrease in amplitude with time during the period 1996--2002.
The temporal variation in the rotation rate and meridional flow should
provide constraints on  dynamo models.
Haber et al.~(2002) have also studied temporal variation in the
meridional flow pattern using ring diagram analysis.
Their results also suggest that there may be reduction in meridional
velocity with activity.
They have claimed to detect
an occurrence of submerged meridional cell in the northern
hemisphere. We do not find this pattern in our results and the cause
of this discrepancy is not clear. Haber et al.~(2002) also
find that the latitudinal gradient of the meridional
flows  near the equator steepens with time, this 
is not clear in our work.
More work is needed to study
possible systematic errors in ring diagram analysis before it can
be used to study small temporal variations reliably. 
In particular, the flow pattern is likely to be modified in the
vicinity of active regions and presence of large number of
active regions during high activity period can modify the
inferences even when average over a large number of regions is
taken to find the mean meridional flow velocity.

While discussing the time variability of the meridional flow
or the north-south antisymmetric component of the zonal flow, one should
keep in mind that any time variations in the instrument properties
will result in a spurious time variation of the flows. This is
less true for symmetric component of zonal flows since global modes are not affected as much.
The zonal flow results are also confirmed by using two different sets of
data (GONG and MDI), while in the case of the meridional flows
we have only used the MDI data. It would be interesting to verify
these results using GONG+ data that is now becoming available.
There is some evidence that some
systematic error may have been introduced in the MDI data during recovery
of the SOHO satellite (Antia 2002). It is not clear how this will affect
the meridional flow measurements.

\section*{ACKNOWLEDGMENTS}

This work  utilizes data obtained by the Global Oscillation
Network Group (GONG) project, managed by the National Solar Observatory,
which is
operated by AURA, Inc. under a cooperative agreement with the
National Science Foundation. The data were acquired by instruments
operated by the Big Bear Solar Observatory, High Altitude Observatory,
Learmonth Solar Observatory, Udaipur Solar Observatory, Instituto de
Astrofisico de Canarias, and Cerro Tololo Inter-American Observatory.
This work also utilizes data from the Solar Oscillations
Investigation/ Michelson Doppler Imager (SOI/MDI) on the Solar
and Heliospheric Observatory (SOHO).  SOHO is a project of
international cooperation between ESA and NASA.
MDI is supported by NASA grants NAG5-8878 and NAG5-10483
to Stanford University.
The authors would like to thank the SOI Science Support Center
and the SOI Ring Diagrams Team for assistance in data
processing. The data-processing modules used were
developed by Luiz A. Discher de Sa and Rick Bogart, with
contributions from Irene Gonz\'alez Hern\'andez and Peter Giles.
This work was supported in part by NASA
Grant NAG5-10912 to SB.

\clearpage

\begin{table}[h]
\caption{Regions studied}
\label{tab:tab1}
\begin{tabular}{lccccc}
\tableline
No.&Carr.&No. of & Lon. & Time & 10.7cm\\
& Rot. & spectra & & & Flux \\
\tableline
1 & 1911 & 12 & 285-120 &1996.07 & $70.5\pm0.5$\\
2 & 1921 & \phantom{0}8 & 120-015 & 1997.04& $71.4\pm0.5$\\
3 & 1932 & \phantom{0}9 & 360-240 & 1998.02& $84.8\pm 1.3$\\
4 & 1948 & 24 & 360-015 & 1999.04& $120.4\pm 2.5$\\
5 & 1964 & 24 & 360-015 & 2000.06& $188.9\pm3.8$\\
6 & $\!\!\!$1975+76$\!\!\!$& 24 & 120-135 & 2001.04-5& $154.1\pm3.2$\\
7 & 1985 & 15 & 300-075  & 2002.01& $224.0\pm3.4$\\
8 & 1988 & 24 & 360-15 & 2002.03-4 & $197.2\pm2.9$\\
\tableline
      \end{tabular}
\end{table}

\clearpage

\begin{table}[h]
\caption{Properties of the tachocline at a few selected latitudes:
All quantities  are the weighted averages of
the results obtained by all three techniques
mentioned in Section 1.2  from all the non-overlapping data sets
from GONG and MDI. 
It is assumed that the different measurements are uncorrelated.}
\smallskip
\label{tab:tab2}
\begin{tabular}{lccc}
\tableline
Latitude&$\delta\Omega_t$&$r_t$&$w$\\
($^\circ$)&(nHz)&($R_\odot$)&($R_\odot$)\\
\tableline
\phantom{0}0&$\phantom{-}20.82\pm0.43$&$0.6916\pm0.0019$&$0.0065\pm0.0013$\cr
15&$\phantom{-}17.83\pm0.24$&$0.6909\pm0.0018$&$0.0078\pm0.0013$\cr
45&$-30.54\pm0.54$&$0.7096\pm0.0019$&$0.0103\pm0.0012$\cr
60&$-67.65\pm0.74$&$0.7104\pm0.0022$&$0.0151\pm0.0020$\cr
\tableline
      \end{tabular}
\end{table}

\clearpage

\clearpage

\clearpage

\clearpage

\clearpage

\clearpage

\clearpage

\clearpage

\clearpage

\clearpage

\clearpage

\clearpage

\clearpage

\clearpage

\begin{thebibliography}{}

\bibitem[Anderson et al.~(1990)]{an90}
Anderson, E. R., Duvall, T. L., Jr., \&
Jefferies, S. M. 1990, ApJ, 364, 699

\bibitem[Antia~(2002)]{ant02}
Antia, H. M. 2002, in Magnetic Coupling of
the Solar Atmosphere, Proc. IAU Coll. 188
(ESA SP-505; Noordwijk: ESA) (in press) (astro-ph/0208339)

\bibitem[Antia \& Basu~(2000)]{ant00}
Antia, H. M., \& Basu, S. 2000, ApJ, 541, 442

\bibitem[Antia \& Basu~(2001)]{an00}%
Antia, H. M., \& Basu, S. 2001, ApJ, 559, L67

\bibitem[ANtia, Basu, \& Chitre~(2001)]{ant98}
Antia, H. M., Basu, S., \& Chitre, S. M. 1998, MNRAS, 298, 543

\bibitem[Antia et al.~(2001)]{ant01}
Antia, H. M., Basu, S., Hill, F., Howe, R., Komm, R. W.,
\& Schou, J. 2001, {MNRAS}, 327, 1029

\bibitem[Basu~(2002)]{ba02}
Basu, S. 2002,  in From Solar Minimum to Maximum: Half a Solar Cycle with SOHO,
Proc. SOHO11 Workshop, ed.~A. Wilson (ESA SP-508; 
Noordwijk: ESA), 7

\bibitem[Basu \& Antia~(1997)]{ba97}
Basu, S., \& Antia, H. M. 1997, MNRAS, 287, 189

\bibitem[Basu \& Antia~(1999)]{ba99}
Basu, S., \& Antia, H. M. 1999, ApJ, 525, 517

\bibitem[Basu \& Antia~(2000a)]{ba00a}
Basu, S., \& Antia, H. M. 2000a, Sol.\ Phys., 192, 449

\bibitem[Basu \& Antia~(2000b)]{ba00b}
Basu, S., \& Antia, H. M. 2000b, Sol.\ Phys., 192, 469

\bibitem[Basu \& Antia~(2001)]{ba01} Basu, S., \& Antia, H. M.
2001, MNRAS, 324, 498

\bibitem[Basu, Antia \& Tripathy~(1999)]{baa99}
Basu, S., Antia, H. M., \& Tripathy, S. C. 1999, ApJ, 512, 458

\bibitem[Benevolenskaya, Kosovichev \& Scherrer~(2001)]{ben01}
Benevolenskaya, E. E., Kosovichev, A. G., \& Scherrer, P. H.
2001, ApJ, 554, L107

\bibitem[Bogart et al.~(1997)]{bog97}
Bogart, R. S., et al.~1997,
in Sounding Solar and Stellar Interiors, IAU Symposium 181,
eds. J. Provost \& F.-X. Schmider, Kluwer Academic Publishers, 111


\bibitem[Bogart, Basu \& Antia~(2002)]{bog02}
Bogart, R. S., Basu, S., \& Antia, H. M. 2002, 
in From Solar Minimum to Maximum: Half a Solar Cycle with SOHO,
Proc. SOHO11 Workshop, ed.~A. Wilson (ESA SP-508; 
Noordwijk: ESA), 145

\bibitem[Brown et al.~(1989)]{br89}
Brown, T. M., Christensen-Dalsgaard, J., Dziembowski, W. A., 
Goode, P., Gough, D. O., \& Morrow, C. A. 1989, ApJ, 343, 526


\bibitem[Brun \& Toomre~(2001)]{bru01}
Brun, A. S., \& Toomre, J. 2001, in Helio- and Astero-seismology at the Dawn of the Millennium,
ed. A. Wilson, (ESA SP-464; Noordwijk: ESA), 619

\bibitem[Brun, Turck-Chi\`eze \& Zahn~(1999)]{bru99} 
Brun, A. S., Turck-Chi\`eze, S., \& Zahn, J. P. 1999,
{ApJ,} {525}, 1032

\bibitem[Charbonneau et al~(1999)]{cha99}
Charbonneau, P., \jcd, J., Henning, R., Larsen, R. M., Schou, J., Thompson, M. J.,
\& Tomczyk, S.  1999, ApJ, 527, 445

\bibitem[Christensen-Dalsgaard \& Schou~(1988)]{jcd88}
\jcd, J., \& Schou, J. 1988, in Seismology of the Sun and Sun-like Stars, 
ed. E.J.Rolfe (ESA SP-286; Noordwijk: ESA), 149

\bibitem[Christensen-Dalsgaard et al.~(1991)]{jcd91} 
\jcd, J., Gough, D. O., \& Thompson, M. J. 1991, ApJ, 378, 413

\bibitem[Christensen-Dalsgaard et al.~(1996)]{jcd96} 
\jcd, J., et al.~1996, Sci., 272, 1286

\bibitem[Corbard et al.~(2001)]{cor01}
Corbard, T., Jim\'enez-Reyes, S. J., Tomczyk, S., Dikpati, M., \&
Gilman, P. 2001, in Helio- and Astero-seismology at the Dawn of
the Millennium, ed. A. Wilson (ESA SP-464; Noordwijk: ESA), 265

\bibitem[Covas et al.~(2000)]{cov00}
Covas, E., Tavakol, R., Moss, D., \& Tworkowski, A. 2000, A\&A, 360, L21

\bibitem[Covas et al.~(2001)]{cov01}
Covas, E., Tavakol, R., \& Moss, D. 2001, A\&A, 371, 718

\bibitem[Dikpati \& Charbonneau~(1999)]{dik99}
Dikpati, M., \& Charbonneau, P. 1999, ApJ, 518, 508

\bibitem[Dikpati \& Gilman~(2001)]{dik01}
Dikpati, M., \& Gilman, P. 2001, ApJ, 559, 428

\bibitem[Elsworth et al>~(1990)]{els90}
Elsworth, Y., Howe, R., Isaak, G. R., McLeod, C. P.,
\& New, R. 1990, Nature, 345, 322

\bibitem[Erofeev \& Erofeeva~(2000)]{ero00}
Erofeev, D. V., \& Erofeeva, A. V. 2000, Sol.\ Phys., 191, 281

\bibitem[Giles et al.~(1997)]{gi97}
Giles, P. M, Duvall, T. L., Jr., Scherrer, P. H., \& Bogart, R. S.
1997, Nature, 390, 52


\bibitem[Haber et al.~(2001)]{hab01}
Haber, D. A., Hindman, B. W., Toomre, J., Bogart, R. S., \& 
Hill, F. 2001, in Helio- and Astero-seismology at the Dawn of the Millennium,
ed. A. Wilson (ESA SP-464; Noordwijk: ESA), 209

\bibitem[Haber et al.~(2002)]{hab02}
Haber, D. A., Hindman, B. W., Toomre, J., Bogart,  R. S., 
Larsen, R. M., \& Hill, F. 2002, ApJ, 570, 855

\bibitem[Hathaway et al.~(1996)]{hat96}
Hathaway, D. H., et al.~1996, Sci., 272, 1306

\bibitem[Hill~(1988)]{hi88}
Hill, F. 1988, ApJ, 333, 996

\bibitem[Hill et al.~(1996)]{hi96}
Hill, F., et al.~1996, Sci., 272, 1292

\bibitem[Howard \& LaBonte~(1980)]{how80} 
Howard, R., \& LaBonte, B. J. 1980, ApJ, 239, L33


\bibitem[Howe et al.~(2000a)]{how00}
Howe, R., Christensen-Dalsgaard, J., Hill, F., Komm, R. W.,
Larsen, R. M., Schou, J., Thompson, M. J., \& Toomre, J. 2000a, ApJ, 533, L163

\bibitem[Howe et al.~(2000b)]{ho00}%
Howe, R., \jcd, J., Hill, F., Komm, R. W.,
Larsen, R. M., Schou, J., Thompson, M. J., \& Toomre, J. 2000b, Sci, 287, 2456

\bibitem[2001a]{howe0} Howe, R., Christensen-Dalsgaard, J., Hill, F., 
Komm, R. W., Larsen, R. M., Schou, J.,  Thompson, M. J., \& Toomre, J. 
2001a, in Helio- and Asteroseismology at the Dawn of the Millennium,
ed. A. Wilson (ESA SP-464; Noordwijk: ESA), 19

\bibitem[Howe et al.~(2001b)]{how01}
Howe, R., Hill, F., Basu, S., Christensen-Dalsgaard, J., Komm, R. W.,
Larsen, R. M., Roth, M., Schou, J., Thompson, M. J., \& Toomre, J.
2001b, in Helio- and Astero-seismology at the Dawn of the Millennium,
ed. A. Wilson (ESA SP-464; Noordwijk: ESA), 137

\bibitem[Jim\'enez-Reyes et al.~(2001)]{jim01}
Jim\'enez-Reyes, S. J., Corbard, T., Pall\'e, P. L., Roca Cort\'es, T.,
\& Tomczyk, S. 2001, A\&A, 379, 622

\bibitem[Komm, Howard \& Harvey~(1993)]{kom93}
Komm, R. W., Howard, R. F., \& Harvey, J. W. 1993, Sol.\ Phys., 143, 19

\bibitem[Korzennik, Rabello-Soares \& Schou~(2002)]{kor02}
Korzennik, S. G., Rabello-Soares, M. C., \& Schou, J. 2002, (astro-ph/0207371)

\bibitem[Kosovichev~(1996)]{kos96}
Kosovichev, A. G. 1996, ApJ, 469, L61

\bibitem[Kosovichev et al.~(1997)]{kos97}
Kosovichev, A. G., et al.~1997, Sol.\ Phys., 170, 43

\bibitem[LaBonte \& Howard~(1982)]{lab82}
LaBonte, B. J., \& Howard, R. 1982, Sol.\ Phys., 75, 161

\bibitem[LeRoy \& Noens~(1983)]{le83}
Leroy, J. -L., \& Noens, J. -C. 1983, A\&A, 120, L1

\bibitem[Libbrecht~(1989)]{lib89}%
Libbrecht, K. G.  1989, ApJ, 226, 1092

\bibitem[Libbrrecht \& Woodard~(1990)]{lib90}
Libbrecht, K. G., \& Woodard, M. F. 1990, Nature, 345, 779

\bibitem[Makarov \& Sivaraman~(1989)]{mak89}
Makarov, V. I., \& Sivaraman, K. R. 1989, Sol.\ Phys., 123, 367

\bibitem[Miesch~(2000)]{mie00}
Miesch, M. 2000, Sol.\ Phys., 192, 59

\bibitem[Nandy \& Choudhuri~(2002)]{nan02}
Nandy, D., \& Choudhuri, A. R. 2002, Sci., 296, 1671

\bibitem[Nigam \& Kosovichev~(1998)]{ni98}
Nigam, R., \& Kosovichev, A. G. 1998, ApJ, 505, L51

\bibitem[Patr\'on et al.~(1997)]{pa97}
Patr\'on, J., et al.~1997,
ApJ, 485, 869

\bibitem[Rajaguru, Basu \& Antia~(2001)]{raj01}
Rajaguru, S. P., Basu, S., \& Antia, H. M. 2001, ApJ, 563, 401

\bibitem[Richard et al.~(1996)]{ric96} Richard, O., Vauclair, S., Charbonnel, C., \&
Dziembowski, W. A. 1996,
{A\&A}, {312}, 1000

\bibitem[Rhodes et al.~(1998)]{rho98}
Rhodes, E. J., Jr., Reiter, J., Kosovichev, A. G., Schou, J., \&
Scherrer, P. H. 1998, in 
Structure and Dynamics of the Interior of the Sun and Sun-like Stars:
Proc.~SOHO 6/GONG 1998 Workshop, ed.~S. Korzennik \& A. Wilson
(ESA SP-418; Noordwijk: ESA), 73

\bibitem[Ritzwoller \& Lavely~(1991)]{rit91}
Ritzwoller, M. H., \& Lavely, E. M. 1991, ApJ, 369, 557

\bibitem[Schou, Christensen-Dalsgaard \& Thompson~(1994)]{sch94} 
Schou, J., Christensen-Dalsgaard, J., \& Thompson, M. J. 1994, ApJ, 433, 389

\bibitem[Schou~(1999)]{sch99}
Schou, J. 1999, ApJ, 523, L181

\bibitem[Schou et al.~(1998)]{sch98}
Schou, J., et al.~1998, ApJ, 505, 390

\bibitem[Schou et al.~(2002)]{sch02}
Schou, J., et al.~2002, ApJ, 567, 1234

\bibitem[Snodgrass~(1984)]{sno84} Snodgrass, H. B. 1984, Solar Phys.,
94, 13

\bibitem[Snodgrass~(1992)]{sno92} Snodgrass, H. B. 1992, in The Solar Cycle,
Proc. NSO 12th Summer Workshop, ASPS 27, 205

\bibitem[Spiegel \& Zahn~(1992)]{spi92}%
Spiegel, E. A., \& Zahn, J.-P. 1992, A\&A, 265, 106

\bibitem[Thompson et al.~(1996)]{tho96}
Thompson, M. J., et al.~1996, Sci., 272, 1300


\bibitem[Toomreet al.(2000)]{too00}
Toomre, J., \jcd, J., Howe, R., Larsen, R. M., Schou, J., \& Thompson, M. J.
2000, Sol.\ Phys., 192, 437

\bibitem[Ulrich~(2001)]{ulr01}
Ulrich, R. K. 2001, ApJ, 560, 466

\bibitem[Ulrich~(1988)]{ulr88}
Ulrich, R. K., Boyden, L. W., Snodgrass, H. B., Padilla, S. P.,
Gilman, P., \& Shieber, T. 1988, Sol.\ Phys., 117, 291

\bibitem[Vorontsov~(2001)]{vor01}
Vorontsov, S. V. 2001, in 
Helio- and Astero-seismology at the dawn of the millennium:
Proc.~SOHO 10/GONG 2000 Workshop, ed.~A. Wilson
(ESA SP-464; Noordwijk: ESA), 563

\bibitem[Vorontsov~(2002)]{vor02}
Vorontsov, S. V. 2002,
in From Solar Minimum to Maximum: Half a Solar Cycle with SOHO,
Proc. SOHO11 Workshop, ed.~A. Wilson (ESA SP-508; 
Noordwijk: ESA), 107

\bibitem[Vorontsov et al.~(2002)]{vetal02}
Vorontsov, S. V., \jcd, J., Schou, J., Strakhov, V. N.,
\& Thompson, M. J. 2002, Sci., 296, 101

\bibitem[Woodard~(2000)]{woo00}
Woodard, M. F. 2000, Sol.\ Phys., 197, 11

\end{thebibliography}
\end{document}